\documentclass[sigconf]{acmart}

\usepackage{balance}

\def\BibTeX{{\rm B\kern-.05em{\sc i\kern-.025em b}\kern-.08emT\kern-.1667em\lower.7ex\hbox{E}\kern-.125emX}}
    
\copyrightyear{2020}
\acmYear{2020}
\setcopyright{acmlicensed}
\acmConference[AIES '20]{Proceedings of the 2020 AAAI/ACM Conference on AI, Ethics, and Society}{February 7--8, 2020}{New York, NY, USA}
\acmBooktitle{Proceedings of the 2020 AAAI/ACM Conference on AI, Ethics, and Society (AIES '20), February 7--8, 2020, New York, NY, USA}
\acmPrice{15.00}
\acmDOI{10.1145/3375627.3375863}
\acmISBN{978-1-4503-7110-0/20/02}

\newtheorem{theorem}{Claim}

\begin{document}

\fancyhead{}

\title{Social and Governance Implications of Improved Data Efficiency}


\author{Aaron D. Tucker}\authornote{Work done as a Summer Fellow at Centre for the Governance of AI}
\affiliation{%
  \institution{Department of Computer Science, Cornell University}}
  \affiliation{%
  \institution{\vspace{2mm} Centre for the Governance of AI, Future of Humanity Institute, University of Oxford}}
\email{aarondtucker@cs.cornell.edu}

\author{Markus Anderljung}
\affiliation{%
  \institution{Centre for the Governance of AI, Future of Humanity Institute, University of Oxford}}
\email{markus.anderljung@governance.ai}

\author{Allan Dafoe}
\affiliation{%
  \institution{Department of Politics and International Relations \\ \vspace{2mm}  Centre for the Governance of AI, Future of Humanity Institute, University of Oxford}}
\email{allan.dafoe@governance.ai}

%
\renewcommand{\shortauthors}{A. Tucker, M. Anderljung, A. Dafoe}

%
\begin{abstract}
Many researchers work on improving the data efficiency of machine learning.
What would happen if they succeed? This paper explores the social-economic impact of increased data efficiency. Specifically, we examine the intuition that data efficiency will erode the barriers to entry protecting incumbent data-rich AI firms, exposing them to more competition from data-poor firms. 
We find that this intuition is only partially correct: data efficiency makes it easier to create ML applications, but large AI firms may have more to gain from higher performing AI systems.
Further, we find that the effect on privacy, data markets, robustness, and misuse are complex. For example, while it seems intuitive that misuse risk would increase along with data efficiency -- as more actors gain access to any level of capability -- the net effect crucially depends on how much defensive measures are improved.
More investigation into data efficiency, as well as research into the ``AI production function", will be key to understanding the development of the AI industry and its societal impacts.
\end{abstract}

%
%
\ccsdesc[500]{Social and professional topics~Economic impact}
\ccsdesc[500]{Computing methodologies~Machine learning}
\ccsdesc[300]{Computing methodologies~Transfer learning}
\ccsdesc[300]{Theory of computation~Active learning}
\ccsdesc[300]{Security and privacy~Economics of security and privacy}
\ccsdesc[300]{Social and professional topics~Surveillance}

%
\keywords{
Data efficiency,
Production function,
Competitive Advantage,
Transfer learning,
Active learning,
Data markets
}

%
\maketitle

\section{Introduction}
How does the performance of an artificial intelligence (AI) system scale with more data, more computational resources, and better algorithms? In other words, what is the \emph{AI production function}\footnote{We depart slightly from the standard definition of production functions \cite{cobbdouglas} by focusing on the relationship between the inputs to a machine learning (ML) system and the performance of the system, rather than between the inputs and outputs of using the ML system on a specific task.}?
This question influences the shape of AI progress, the structure of the AI industry, and the societal impacts of AI.

In this paper, we offer a preliminary analysis of one aspect of the AI production function - data. Specifically, we analyze the implications of increases in data efficiency\footnote{Computer scientists may recognize this as being related to sample complexity. We prefer the term data efficiency because it is more intuitive to map "more efficient" to "higher performance", than "lower complexity" to "higher performance". Further, we do not mean to imply any statistical properties of our data (in contrast to the word "sample").}: increases in the performance a system achieves for any given data input. We find that increases in data efficiency have an \textit{access effect} -- where more actors get access to ML capabilities -- and a \textit{performance effect} -- where performance for any given dataset is improved.

\begin{figure}
  \caption{The Access Effect}
  \centering
    \includegraphics[width=0.3\textwidth]{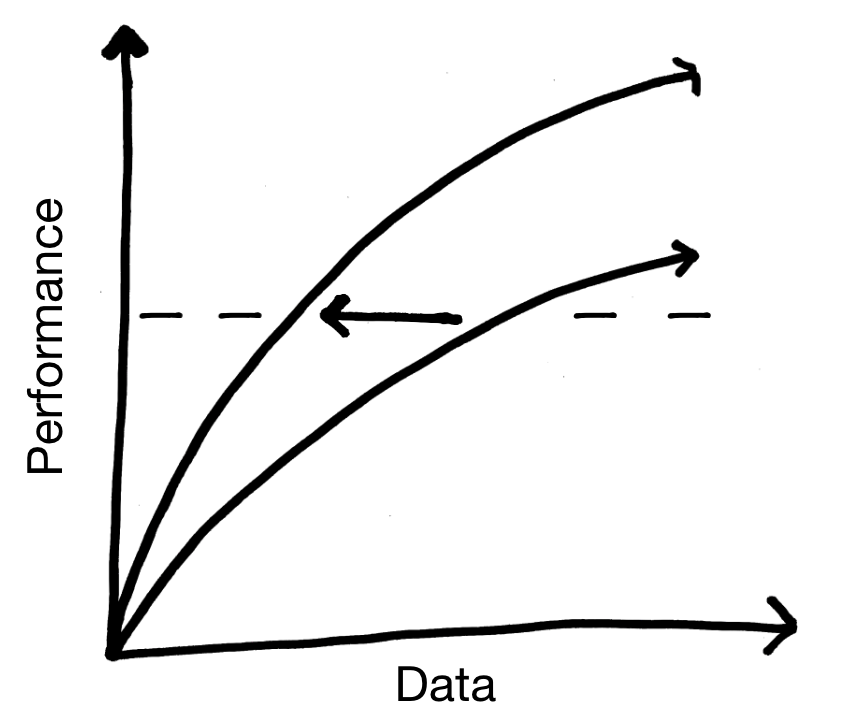}
\end{figure}
\begin{figure}
  \caption{The Performance Effect}
  \centering
    \includegraphics[width=0.3\textwidth]{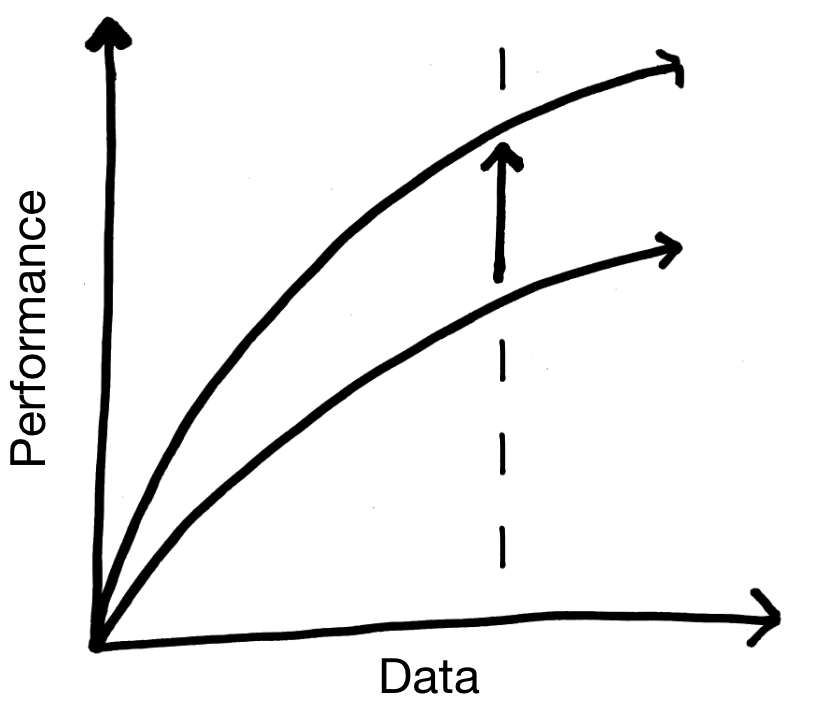}
\end{figure}

Predictions about the AI industry and its societal impacts often implicitly rely on claims regarding the AI production function and specifically on the relationship between data and the performance of systems.
Kai-Fu Lee, for example, claims that China has an advantage with regards to developing AI technology, partly because we live in an ``Age of Data", where ``once computing power and engineering talent reach a certain threshold, the quantity of data becomes decisive in determining the ... accuracy of an algorithm" \cite{kaifulee}. Views on the AI production function – in particular current methods' data efficiency – often also inform views about the limits of machine learning, with many leading researchers for example Yann LeCun, Geoffrey Hinton, Yoshua Bengio (interviewed in \cite{architects}), and Gary Marcus \cite{marcus}, commenting that the need for large amounts of data suggest limitations of our current techniques.

Data efficiency is likely to increase. Many recent advances in Machine Learning and Artificial Intelligence have come from deep learning, a technique which enables high performance on challenging tasks such as image recognition at the cost of large compute requirements as well as needing large data sets. At the same time, we know that improvements to data efficiency are possible. For instance, humans can learn simple visual concepts such as novel characters from single instances \cite{oneshot}. In addition, researchers are interested in making more efficient machine learning algorithms, which decrease the amount of data or compute necessary to achieve some level of performance. For example, transfer learning seeks to use pretraining on one dataset to improve performance on another problem or dataset \cite{transferlearning}\footnote{We include transfer learning as a method for data efficiency because it improves the data efficiency of the target task.}. Few and zero-shot learning seek to successfully learn classifications from very few examples \cite{zeroshot}. Active learning seeks to increase the value of each data point by getting more informative data points \cite{activelearning}. Data efficiency can also be improved by improving data quality, for instance by developing better sensors.

This paper summarises the empirical literature on how performance scales with data, and introduces a simple model to analyze the implications of improved data efficiency. We then explore implications for the AI industry and society, followed by suggestions for directions for future work. 

\section{Prior Work on the Data Efficiency and Production Function of AI}

\subsubsection{Compute costs over time} Recent authors have estimated that the amount of compute used in the largest AI training runs is increasing exponentially, doubling every 3.5 months \cite{aiandcompute}. Some authors suggest this is reason for skepticism about future AI progress, since requiring exponentially more resources to achieve results will become prohibitively expensive \cite{interpretingaiandcompute, reinterpretingaiandcompute}.
These high compute requirements, at least in reinforcement learning applications where the compute is being used to produce training data, suggest there would be large gains from improvements in data efficiency.

\subsubsection{The relationship between performance and data} Recent empirical work suggests that there may be a simple relationship between data, model size, and performance.

One investigation found that across different settings, a power law model (where performance is proportional to the amount of data to some power) described the relationship between the amount of data and performance, as long as the model size grew at a rate dictated by a separate power law \cite{hestness}. If true, these power law relationships allow one to model the data and hardware requirements for a specific performance level, as attempted in recent work \cite{beyond}.

Work by other groups have also found that performance increases with more data, as long as model size is also allowed to increase. While many authors agree on the power law model (for instance \cite{sala}, \cite{mri}, \cite{howmuch}, and \cite{hestness}), other authors find that a logarithmic model explains the relationship \cite{log}. Not all work agrees that there is a simple relationship -- one experiment which did not increase model size found that performance was only marginally improved by increasing the dataset size \cite{contra}.

Note that the overall shape of our data to performance schematics agree with the empirical results of a recent paper investigating data efficient supervised learning \cite{deepminddataefficient}.

\subsubsection{ML performance over time} While there has been excellent work in tracking changes to ML performance over time (for example \cite{aiindex} and \cite{eff}), to our knowledge there are no similar compilations in tracking how data efficiency has changed over time. We believe that this would be a promising direction for future research.

\section{Modeling Data Efficiency: Assumptions}
What are the effects of increasing data efficiency? We construct a simple model of what it means for data efficiency to increase. We model data efficiency as a certain data to performance function, making two assumptions.

Let the \emph{data to performance curve} for an ML system be a function $f$, which takes in a quantity of data $d$, and returns the performance $p$ of the system given that amount of training data. Assume that $f$ is defined in a manner where a larger $p$ is higher performance.

For clarity of presentation, our assumptions tend to be stronger than are necessary for our argument.

\noindent \subsubsection{Assumption 1: Monotonic performance increases}
Adding more data will not decrease system performance according to its performance function – performance will remain increase. We omit considerations of computational cost from our analysis. Formally:

\begin{displaymath}
\forall d, d': d' > d \implies f(d') > f(d)
\end{displaymath}
\begin{center}
``More data always improves performance"
\end{center}

\subsubsection{Assumption 2: Eventually diminishing marginal returns}
Eventually the system will reach a point (here denoted as $m$) where it sees diminishing marginal returns to performance from data. Intuitively, the first time you see something is more informative than the millionth time. This claim has been theoretically shown for some performance functions \cite{learningcurves}.

\begin{displaymath}
\exists m\text{ such that }\forall d>m, d'>m, \Delta > 0 \text{ the following holds}
\end{displaymath}
\begin{displaymath}
d' > d \implies f(d' + \Delta) - f(d') < f(d + \Delta) - f(d)
\end{displaymath}
\begin{center}
``At some point, more data does not help as much as before"
\end{center}

\section{Models of Data Efficiency}
How can we model increased data efficiency as a transformation of a data to performance function: $f$ to a new $f_\text{efficient}$? We discuss three models of data efficiency to demonstrate different intuitions.

\subsubsection{Data efficiency modeled as adding data} Data efficiency can be simply modeled as analogous to giving all users more data.  This model may be appropriate for understanding the impact of transfer learning, where you use data from one source to improve performance on a variety of tasks. Similarly, this model may be appropriate when data efficiency with respect to real world data comes from using additional simulated data. We do not claim that each data point of simulated or transfer data is as useful as a data point for the target task, but rather that they may be equivalent to some amount of data for the target task.

\begin{displaymath}
f_\text{efficient}(d) = f(d + c) \text{ where } c > 0
\end{displaymath}
\begin{center}
``Data efficiency is like adding more data"
\end{center}

\subsubsection{Data efficiency by increasing data value} Another way to model data efficiency is as an increase in the value of data.
This could be accomplished through "better data", for instance by collecting data from better placed sensors, which better capture the phenomena one is trying to model. Another plausible path to increased data value is by using active learning, where each data point is chosen to be more informative to the system.

\begin{displaymath}
f_\text{efficient}(d) = f(a * d) \text{ where } a > 1
\end{displaymath}
\begin{center}
``Data efficiency is like accessing a constant factor more data"
\end{center}

\subsubsection{Data efficiency modeled by function composition} A general formal expression of data efficiency is as follows, where $g$ is monotonic and continuous:
\begin{displaymath}
f_\text{efficient}(d) = f(g(d)) \text{ where } g(d) > d
\end{displaymath}

\subsection{Two core effects of Data Efficiency}
We conceptualize the impacts of data efficiency as composed of two effects -- an access effect and a performance effect. The access effect refers to how any given ML capability becomes more accessible to more actors: a given level of performance becomes accessible with less data. The performance effect refers to how for any given amount of data, it becomes possible to achieve higher performance.

\subsubsection{The Access Effect}

As depicted in Figure 1, the access effect refers to the leftward shift of the data to performance curve. This captures most of the straightforward impacts of improved data efficiency, namely decreased data requirements to achieve any given level of performance. This has the effect of enabling new applications in data limited domains and broadening access of existing capabilities to more actors.

\begin{theorem}Improved data efficiency makes any given level of performance attainable with less data\end{theorem}
\noindent Formally, all of our models of data efficiency transform $d$ in some way such that $g(d) > d$ for every $d$. Assuming continuity, this means that there is some $d' < d$ such that $g(d') \geq d$, and therefore $f_\text{efficient} = f(g(d')) \geq f(d)$ attaining or exceeding the same level of performance with less data.

\subsubsection{The Performance Effect}

As depicted in Figure 2, the performance effect refers to the upward shift in performance for a given amount of data. It can be muted by the presence of performance ceilings or diminishing marginal returns. This increases the level of performance for many levels of data, assuming access to the same algorithms.

\begin{theorem}Improved data efficiency increases performance\end{theorem}

\noindent Formally, all of our models of data efficiency transform $d$ in some way such that $g(d) > d$ for every $d$. By the monotonicity assumption, this fact implies that $f_\text{efficient}(d) = f(g(d)) > f(d)$.

\section{Consequences of Increased Data Efficiency}

\subsection{Impact on ML-based Capabilities and Applications}

\subsubsection{New applications in data limited domains}
One of the clearest implications of data efficiency is the ability to use ML to solve problems in data-limited domains. Data may be limited because there are fundamental limitations -- e.g. the data does not exist -- or because collecting it is expensive.

One notable example of an area where there is a limited amount of obtainable data is ancient languages, where only so many known text fragments exist. Improvements in data efficiency may improve machine translation applications in these domains.

There are many domains where obtaining new data is costly. This can be data from expensive medical or chemical tests, sensors,  real world experiments, or human feedback. As data efficiency improves one would expect ML applications in these domains to become more feasible. This is also the case in domains where data is not presently being collected but potentially could be (for instance, expert judgment for a specific area in a standardized format on difficult questions e.g. medical diagnosis).

A particularly important example for this trend is robotics. Collecting data from real world robots may be expensive because of the costs of robot time (maintenance, damage risk, needing to reset the task, etc.). Relatively recent improvements to data efficiency have made deep reinforcement learning from only real world data possible on simple robotic tasks \cite{sacrobots}\footnote{This is in contrast to methods which are data-efficient with respect to real-world interaction, but which rely on large amounts of simulated data \cite{openaihand}. Those methods are less computationally efficient, and require upfront investment in simulation capabilities.}.

\subsubsection{New actors access ML capabilities}
Another implication of the access effect is that more actors have access to ML capabilities, since one needs less data in order to achieve a given level of performance. This benefits data-poor actors, suggesting that more companies will be able to deliver a (potentially new) product with a certain level of performance.

An interesting type of data-poor actor is a team within a larger organization, which would like access to more resources (e.g. data, compute, or engineers) to develop some ML capability. As data efficiency improves, less data is required to develop a prototype ML application in order to demonstrate the potential application's value. This may smooth out the adoption of ML by organizations  – e.g. government agencies – who have enough data, but lack organizational buy-in to develop applications.

\subsubsection{Misuse potential}
By increasing the number of actors with access to a given ML capability, the chance increases that an actor with malicious ends will also gain access. Researchers have explored the many ways that ML-based applications could be misused, including for cyberattacks, surveillance, and attempts to affect elections \cite{misuse}.

Deepfakes and synthetic media are a popular example of technologies with a high potential for misuse. In fact, the risk from these systems can be understood as a product of their high data efficiency. Deepfakes that required 1000s of hours of video would be much less disruptive, whereas a recent system was able to base deepfakes on as few as 32 video frames \cite{fewshotdeepfake}. Data efficiency is a crucial parameter in judging misuse potential.

However, the net effect on misuse from increased data efficiency is complex. Many malicious uses can be defended against. Therefore, as data efficiency increases, we can also expect more actors to gain access to defensive capacities as well as the development of more powerful defenses. The net effect will therefore depend e.g. on the offense-defence balance in the relevant domain \cite{odb}, the adoption rates of defensive measures, and the extent to which defender or attacker capabilities scale faster as data efficiency increases. Take the example of cybersecurity. Automated vulnerability detection can be used offensively, but it can also be used defensively in order to pre-emptively detect and patch vulnerabilities prior to releasing systems. The factors regarding the net effect of more actors having access to a technology are generally complex, and vary considerably based on the domain \cite{shevlane}.

\subsection{Impacts on Competitive Advantage}
Competitive advantage is a core concept in Economics, which refers to factors which allow firms to outperform competitors, charge more, offer better services, etc. \cite{competitiveadvantage}. The concentration of AI-related industries and capabilities is an important factor in the governance landscape of AI -- there are different policy options in highly concentrated or highly decentralized situations. Further, technological changes can impact competitive advantage \cite{techcompetitiveadvantage}.

How would improvements to data efficiency affect the competitive advantage of large AI firms? Prima facie, it seems that improvements in data efficiency would lead to a levelling effect, decreasing market concentration. Firstly, the access effect gives more actors access to any level of performance. Secondly, assuming that there are performance bounds to a task, such as for example in facial recognition, the performance effect will eventually diminish the absolute difference in performance between actors. While the above effects may dominate, the overall effect on competitive advantage may in fact be to benefit data-rich actors more than data-poor actors. This is because the value derived from a certain level of performance on a task – say revenue created by a recommendation algorithm – differs greatly between actors and often correlates with the actor's size, and because revenue does not scale linearly with performance.

\subsubsection{Actors derive different amounts of value from the same level of ML performance}  Actors derive different amounts of value from the same performance on a task, and so the performance effect benefits some actors more than others. The value an actor derives from a certain capability depends on access to complements to the technology: e.g. having products to sell, customers to sell those products to, and market access. An ML capability which increases user engagement by a fixed $5\%$ will increase total engagement, revenue, and profit more for actors with large user bases.
 
Furthermore, one can expect being an AI incumbent to correlate with having substantial complements to AI technology. Many contemporary data-rich actors made their investments in ML on the basis of already having more complements to ML performance than other actors. For example, digital advertising may be a domain in which having better ML applications is especially useful, and so companies with large digital advertising revenue may be more inclined to invest in ML. In sum, actors who have more complements to AI technology benefit more from across-the-board increases in performance (such as from improvements to data efficiency) and data rich AI incumbents are likely to have more AI complements, potentially increasing their competitive advantage.

\subsubsection{Winner-take-all markets} Economists often characterize aspects of the AI industry as a winner-take-all, or winner-take-most market. In such a market, what matters most is whether a firm is first, or not; it doesn't matter much for their market share how good their service is in an absolute sense. Since data efficiency does not alter the rank ordering of actors in the performance of their ML systems, holding datasets and other assets constant, it will have no impact on a pure winner-take-all market assuming that all actors have access to the same algorithms. 

\subsubsection{Threshold effects} There are many tasks where a certain threshold of performance is needed before the service has value. Autonomous vehicles, for example, will only become a viable mass consumer product once they exceed some performance threshold. Once this threshold is reached, a company will see a large spike in the value they can reap from the capability. Improvements in data efficiency may therefore lead to increased concentration if it pushes only a small number of actors above the threshold at which a product becomes viable or a task becomes solvable. 

\subsubsection{Potential value near performance ceilings}

Many real-world problems have performance ceilings. For instance, a mean squared error cost function used to measure the performance of an image recognition algorithm has a fundamental performance ceiling at $0$ error. Predicting the outcome of a fair coin has an irreducible error of $50\%$. While it may seem that this would mute competitive advantage stemming from the performance effect, it is not so straightforward. Improved performance can be valuable even close to a performance boundary, and thus a smaller absolute improvement for a data-rich actor may still yield more value than a larger absolute performance improvement for a data-poor actor.

Firstly, having a very high performing system allows one to use its output as the input to other systems. For example, Alipay's Smile to Pay allows users to authenticate payments with their face and access to their mobile phone, showing high confidence in the accuracy of the underlying facial recognition systems \cite{face}.

Secondly, a nominal bound on the performance function does not necessarily imply a practical bound on performance; further, often the marginal benefits of improvement increase as we approach a nominal performance bound.  Consider a hypothetical task with the performance function of $P(\text{task is performed correctly}) = p$. This performance function is trivially bounded by $0$ and $1$. One might think that moving from $P(\text{task is performed correctly}) = 0.99$ to $0.999$ is relatively unimportant. However, the expected number of times that we can perform the task before encountering a single error is simply the negative binomial distribution $\mathcal{NB}(p, 1)$, which has the following expected value. $$\text{Expected number of tasks before error} = \dfrac{p}{1-p}$$

Going from $p = 0.99$ to $p = 0.999$ takes the number of times that we can do the task before an error from $99$ to $999$, almost a 10x improvement. Further, this remains true all the way to the trivial upper bound of $1$ -- going from $0.999$ to $0.9999$ is almost another 10x increase in the expected number of task attempts before error.

In many domains (such as capital investments, survival analysis, etc.), the time until error is the important parameter, rather than the probability of failure in any given unit time. This shows that an apparent bound on the performance function is not necessarily a bound on the utility function.

\subsection{Consequences for Safety and Robustness}
\subsubsection{Distributional shift}
In a much more data-efficient world, high performance is attainable with  access to much less data. This may mean that deployed systems are more sensitive to distributional shifts, since they may be trained on less representative data and because actors will be more tempted to deploy high-performing systems.

Typically, ML practitioners evaluate their models before deployment using the data that they have access to. If performance is good enough, they may choose to deploy the model. Depending on how the dataset is constructed, larger datasets are more likely to contain representatives of relatively unlikely inputs. This means that needing a large dataset to get the necessary level of performance could give some more robustness to distributional shift, if only because it provides more examples, and a better sense of the rarer parts of the distribution. As such, increased data efficiency may increase issues related to distributional shift.

To counteract this effect, developers  ought to think carefully about evaluation and dataset collection – if high performance is possible with a smaller dataset, then it is important to proactively include less well-represented inputs in the evaluation, since they will not be sampled as often.

A similar point is that if ML seemingly works on more problems, then this will increase the extent to which such systems are deployed. If deployment of systems based on smaller training datasets happens before researchers address issues with generalization, then this may lead more people to deploy non-robust ML systems.

\subsubsection{Human oversight}
Human oversight and feedback is a particularly costly type of data. Some methods for AI safety are based on the idea of scalable oversight \cite{concreteproblems}, where one either directs human oversight to be more effective \cite{withouterror}, itself a form of data efficiency, or trains a model of human approval/disapproval and uses that as a safety component in other parts of the system \cite{rewardmodeling}. In a more data efficient world, these methods are more viable.

\subsection{Impact of Marginal vs. Total Value of Data}
Data efficiency, whether modeled as adding data or as increasing data value, both show a performance effect regarding the \textit{total} performance value of data, but they disagree on the \textit{marginal} performance value of data. The additive model yields a lower marginal performance value of data, because of the diminishing marginal returns assumption. The multiplicative model yields a higher marginal performance value of data because of the chain rule. Thus, the effect of increased data efficiency on marginal value of data is an open question according to these models.

\subsubsection{Data Markets} Data markets would likely be greatly affected by changes in data efficiency. Firstly, if the marginal value of data goes up, this may increase actors' willingness to buy, sell, or protect their data. Secondly, the collection of new forms of data may become viable if the marginal value of that data surpasses the marginal cost of collecting it. If the marginal value of data is already greater than the cost of collection and then increases, this may instead be realized as increased profit for data-selling firms, rather than increased data collection.

\subsubsection{Data Labeling} If the marginal value of data increases, then actors will be more willing to pay for data labeling, and there is more potential for higher wage data-labeling jobs, especially where the labeling task is more skill or knowledge intensive. For example, it may become viable to have highly paid professionals such as doctors or lawyers to label data. As ML becomes viable for more tasks, the range of labeling tasks may also expand.

\subsubsection{Surveillance and Privacy}
If the marginal value of data increases this may potentially exacerbate issues in surveillance and privacy. This can potentially be mitigated by the fact that the increased marginal value of data makes it more worthwhile to undergo the expense to collect or process it in a more privacy-preserving manner.

If data efficiency improves in such a way that the marginal value of data decreases, one may expect less surveillance on the margin. However, one might still see a net negative impact on privacy. Firstly, there are likely high fixed costs of building a data collection infrastructure, such that a decrease in the marginal value of data discourages future investments in surveillance, but does not necessarily affect existing data collection infrastructure. Secondly, the performance effect would mean that systems are higher performing overall. As such, the data that the actor already has on its users provides more information about them. An actor would need less data to e.g. predict whether a user is pregnant. As such each piece of data could become arguably more privacy infringing. Further, even if actors are less willing to spend to get new data, as the total value of data increases, they may be more strongly incentivised to hold on to data they have, rather than for instance acquiescing to requests to delete it.

\section{Future Work}

\subsubsection{Data to performance curves} The questions of how performance scales with data using current algorithms, how this has changed over time, and how it is likely to change in the future remain fairly unexplored. Researchers with an interest in these issues can consider conducting empirical tests of the relationships between data and performance, as well as investigations into how algorithmic improvements have affected the data to performance curve.

\subsubsection{Performance to utility functions} A major complicating factor in our analysis is the distinction between performance according to a cost or performance function, and the value provided to a system's owner. What is the owner's utility, for any given level of performance? Research in this area could help yield a more granular and detailed view of the dynamics of AI development, but would require detailed investigation into how exactly ML is used by various actors.

\subsubsection{Production function of AI} The AI production function plays a crucial role in determining parameters relevant to AI governance and ethics, but there are many questions remaining about how the production function of AI works, what to include, how it changes over time, etc. Research in this area would shed significant light on questions of the requirements of AI research, and provide a better understanding of AI progress. 

\subsubsection{Implications of the AI production function} This paper analyzed the potential impact of changes to data efficiency. What would happen if the performance gains of extra computational resources, the size of model, AI talent, or access to state-of-the-art algorithms were to change? Authors may also be interested in studying the extent to which the current structure of the AI industry depends on the AI production function. For example, many  recent  state-of-the-art  results  have come out of private AI labs rather than  universities,  with many researchers moving from university positions into private industry \cite{gofman}. To what extent are these changes a result of e.g. large AI firms having access to large amounts of computational resources and data?

%
\begin{acks}
Thank you to the whole GovAI team from Summer 2019 for lots of useful feedback and fruitful discussions, especially (and in no particular order) Jade Leung, Ben Garfinkel, Jeff Ding, Sören Mindermann,  Ulrike Franke, Maximilian Negele, Max Daniel, Carina Prunkl, Chris Byrd, and Toby Shevlane. Thank you also to other people at FHI who helped clarify many of the ideas in this paper, particularly Eric Drexler, Ryan Carey, and Owain Evans. Thanks to Philip Trammell for some earlier feedback on this paper, and three anonymous reviewers with helpful suggestions. We thank the Open Philanthropy Project for their support of this project. 
\end{acks}

%
\bibliographystyle{ACM-Reference-Format}
\bibliography{bibliography}

\end{document}